\documentclass[twocolumn,showpacs]{revtex4}

\usepackage{graphicx}
\usepackage{hyperref}
\usepackage{xcolor}

\usepackage{mathtools}
\usepackage{times}
\usepackage{latexsym}
\usepackage{graphicx}
\usepackage{amsmath,amssymb}
\usepackage{verbatim,times,bbm}
\usepackage{placeins}
\usepackage{color}
\usepackage[english]{babel}
\usepackage[T1]{fontenc}
\usepackage{amsmath, amsthm, amssymb, mathrsfs}
\usepackage{epstopdf}
\usepackage{subfigure}
\usepackage{appendix}
\usepackage{pdfpages}

\usepackage{placeins}
\usepackage{float}

\usepackage{times}
\usepackage{latexsym}
\usepackage{graphicx}
\usepackage{verbatim,times,bbm}
\usepackage{color}
\usepackage{appendix}

\newcommand{\be}{\begin{equation}}
\newcommand{\ee}{\end{equation}}

\newcommand{\bea}{\begin{eqnarray}}
\newcommand{\eea}{\end{eqnarray}}

\newcommand{\ba}{\begin{array}}
\newcommand{\ea}{\end{array}}

\newcommand{\bn}{\begin{enumerate}}
\newcommand{\en}{\end{enumerate}}

\usepackage{color}

\begin{document}

\title{Persistent homology  analysis of multiqubit entanglement }

\author{Riccardo Mengoni\footnote{email: riccardo.mengoni@univr.it}}
\affiliation{Department of Informatics, University fo Verona, Strada Le Grazie 15, 37134 Verona, Italy}

\author{Alessandra Di Pierro\footnote{email: alessandra.dipierro@univr.it}}
\affiliation{Department of Informatics, University fo Verona, Strada Le Grazie 15, 37134 Verona, Italy}

\author{Laleh Memarzadeh\footnote{email: memarzadeh@sharif.edu}}
\affiliation{Department of Physics, Sharif University of Technology, Tehran, Iran}

\author{Stefano Mancini\footnote{email: stefano.mancini@unicam.it}}
\affiliation{School of Science and Technology, University of Camerino, I-62032 Camerino, Italy}
\affiliation{INFN-Sezione di Perugia, I-06123 Perugia, Italy}

\begin{abstract}
We introduce a homology-based technique for the analysis of multiqubit state vectors. In our approach, we associate  state vectors to 
data sets by introducing a metric-like measure in terms of bipartite entanglement, and
investigate the persistence of homologies at different scales. This leads to a novel  classification of multiqubit entanglement. The relative occurrence frequency of various classes of entangled states is also shown. 
\end{abstract}


\maketitle

\section{Introduction}

Entanglement has been recognized as a key resource for obtaining a quantum boost in many information technology tasks (see e.g. \cite{WGE16}).
As such it deserves a careful characterization. Initial work on the classification of entangled states was focussed on the quantification through so-called `entanglement monotones', i.e. functions of multipartite states that do not increase under local transformations \cite{vidal}. These functions have been proved to work well mostly for bipartite states \cite{Horo}, while for multipartite entanglement another approach seems to be more promising, which is based on partitioning states  according to some notion of equivalence.
In the SLOCC approach \cite{dur}, equivalence classes are constructed on the base of invariance under stochastic local operations and classical communication. However, this leads to infinite (even uncountable) classes for more than three qubit systems. Hence this approach is not effective in the general case, although some ways out were devised for the case of four qubits \cite{four}.

An alternative route to entanglement classification is represented by the analysis of topological features of multipartite quantum states \cite{Quinta,PH1}.
Topological data analysis has recently gained a lot of attention in the classical framework thanks to its suitability for the analysis of huge data sets represented in the form of point clouds: in such cases, it would indeed be impossible to accurately analyze the data, while a ``qualitative" analysis would be efficient. 
Among these techniques, Persistent Homologies (PH) played a pivotal role \cite{Edel,Carl}.
It is a particular sampling-based technique from
algebraic topology aiming at extracting topological 
information from high-dimensional data sets.

Here, following up the work in \cite{PH1}, we apply PH techniques to analyse multiqubit state vectors. Each state vector will be intended as a data set by introducing a metric-like measure in terms of bipartite entanglement. Then the persistence of homologies at different scales will be investigated. 
While the aim of \cite{PH1} was the classification of all possible states for 3 and 4 qubit,
here we focus on `genuine entanglement' and show the classification up to 6 qubit.
Moreover, in this paper we  will also compute the relative occurrence frequency of the various classes of entangled states, by means of a random generation of states.

The article is organized as follow. In Section~\ref{perho} we briefly recall concepts of algebraic topology that will be used thereafter.
In Section~\ref{create} we illustrate the method to produce a data cloud from multiqubit states.
Then we produce and show the barcodes for the cases of 4 and 5 qubit in Section~\ref{classification}
(those for 6 qubit are reported in the Appendix). 
There we also analyze the occurrence frequency of various classes of entangled states.
Finally, we draw our conclusions in Section~\ref{conclusion}.


\section{Persistent Homology}
\label{perho}

A data  cloud is a collection of points in some  $n$-dimensional space $ \mathbb{S}_n$. In many cases, analysing the global 'shape' of the point cloud gives essential insights about the problem it represents. In the context of data analysis, Persistent Homology is an algebraic method for computing coarse topological features of a given data cloud  that persist over many  grouping scales. In this section we review the mathematical background that is necessary to understand this technique \cite{Hatcher,Edelsbrunner}.

Consider  a data cloud  represented by a  set of points  $ \left\lbrace x_\alpha \right\rbrace  $ living  in a Euclidean space. Choosing  a value of the grouping scale $ \epsilon $, it is possible to construct the graph whose  vertices are the data points $ \left\lbrace {x_\alpha} \right\rbrace  $ and edges   $ e_{x_\alpha,x_{\alpha'}}  $ are  drawn when the $ \frac{\epsilon}{2}$-balls  centered in the vertices   ${x_\alpha} $ and $x_{\alpha'}$ intersect each other. Such  graphs show  connected components and hence  clusters obtained   at    $ \epsilon $ scale   but  do not provide information about  higher-order features such as holes and voids.  In order to track  high-dimensional features we need to introduce the following concepts.

\medskip

\textit{Convex set. } A convex set is a region of a Euclidean space where every two points are connected by a straight line segment that is also within the region.

\medskip

\textit{Convex Hull.} The convex hull of a set $ X $ of points in an Euclidean space is the smallest convex set that contains $ X $.

\medskip

\textit{$k$-Simplex.} A $k$-simplex  is a $k$-dimensional polytope which is the convex hull of its $k+1$ vertices. 
Thus, for example, simplices of dimension $ 0, 1, 2 $ and $ 3 $ are respectively vertices, edges, triangles and tetrahedra.

\medskip

\textit{Simplicial Complex.} A simplicial complex $ {\mathcal {K}} $ is a set of simplices that satisfies the conditions:
\begin{enumerate}
	\item[i)] Any face of a simplex from $  {\mathcal {K}} $ is also in  $ {\mathcal {K}} $.
	\item[ii)] The intersection of any two simplices $  \sigma_{1},\sigma_{2}\in {\mathcal {K}} $ is either the empty set   $ \emptyset $  or a face of both  $ \sigma_{1} $ and  $ \sigma_{2} $.
\end{enumerate}
The dimension of a simplicial complex  is equal to the largest dimension of its simplices.

\medskip

\paragraph*{Homology of a complex.}
 For each simplicial complex $ {\mathcal {K}} $ there is a set of homological groups $\left\lbrace H_{0}({\mathcal {K}}),H_{1}({\mathcal {K}}),H_{2}({\mathcal {K}}),\ldots \right\rbrace$, where the $k$th homology group $H_k({\mathcal {K}})$ is non-empty when the $k$-dimensional holes are in ${\mathcal {K}}$. Hence, the homological groups of the simplicial complex describe the order of the holes existing in that simplicial complex. 

\medskip

In order to recognize global topological features of a data cloud it is necessary to  complete  the corresponding graph  to a simplicial complex by filling in the graph with all the  simplices.
Given a grouping scale  $ \epsilon $, there are different   methods to generate  simplicial complexes. In this paper we will focus on the  Rips  complex, $ \mathcal{R}_\epsilon$, where
$k$-simplices  correspond to $(k + 1)$ points  which are pairwise within distance $ \epsilon $.

Topological features of the data cloud are obtained by constructing a homology of the simplicial complex. The homology of the  Rips complex  hence  reveal those topological features  that appear at a chosen value of  $ \epsilon $.
If  $ \epsilon $ is taken too small, then only  multiple connected components are shown. On the other hand, when $ \epsilon $ is  large,  any pairs of points get connected and a giant simplex with trivial homology is obtained. However, it is preferable to make the whole process independent from  the choice of $\epsilon$. 
In order to obtain significant features it is necessary to consider all  the range of   $ \epsilon $. Those topological features
which persist over a significant  interval of the parameter $ \epsilon$ are to be considered  specific of that point cloud, while  short-lived features as less important ones.  
Consider the sequence of Rips complexes $\mathsf{R}=\left\lbrace  \mathcal{R}_{\epsilon_i} \right\rbrace_{i=1}^N $ associated to a given point cloud;
instead of examining the homology of the individual terms $ H(\mathcal{R}_{\epsilon_i}) $, we look at the  inclusion maps  $ I  : H(\mathcal{R}_{\epsilon_i}) \rightarrow H(\mathcal{R}_{\epsilon_j})$ for all $ i < j $. These maps are able to tell us  which features persist since they reveal information that is not visible if we consider $ H(\mathcal{R}_{\epsilon_i}) $ and $ H(\mathcal{R}_{\epsilon_j}) $ separately. 

\medskip

\textit{Barcodes.} Given  the sequence of Rips complexes $\mathsf{R}=\left\lbrace  \mathcal{R}_{\epsilon_i} \right\rbrace_{i=1}^N $, a \textit{barcode}  is a graphical representation of $ H_k(\mathsf{R}) $ as a collection of horizontal line segments in a plane whose horizontal axis corresponds to the parameter $ \epsilon $ and whose vertical axis represents an (arbitrary) ordering of homology groups.  A barcode can be seen as a variation over $ \epsilon $ of the Betti numbers which count the number of $n$-dimensional holes on a simplicial complex  $\mathcal{R}_{\epsilon}$ (cf.  \cite{Barcodes}).


\section{Creating qubit data cloud}
\label{create}

In this section we discuss the methodology we use for creating the data cloud which will be at basis of  classifying entangled states.
We will restrict our attention to $N$ qubit states showing "genuine" entanglement, i.e. that are $N$-partite entangled or ``fully inseparable".

Our approach starts with the random generation of pure states among which we select, using generalised concurrence measure,  those  showing  genuine entanglement.
At this stage, a data cloud is associated  to a state in such a way that each qubit is identified with  a single point in the cloud,  while  a distance between pairs of points   is defined using a semi-metric  that  takes into account  the pairwise entanglement shared by  the two qubits that the  points represent.     Semi-distances between qubits  are stored in a matrix $ D $ which will be the input of the persistent homology algorithm. 

Note that in the definitions given in Section \ref{perho} we refer for simplicity  to Euclidean spaces. Since here we are dealing instead with  a  semi-metric space, it is worth stressing  that  computing   persistent homology is  still possible in our case. In fact, a  distance  between   pairs of points which does not  satisfy triangular inequality is still sufficient     for constructing    Rips complexes.

\subsection{Random state generation}

In order to  randomly generate  a pure    state of $ N $ qubits,  we   employ the following parametrization \cite{Zyczkowski_2001, PhysRevA.94.022341}
\begin{equation}
|\psi\rangle=\sum_{n=0}^{2^N-1} \nu_n |n\rangle,	
\end{equation}
with 
\begin{eqnarray}
\nu_0&=&\cos\theta_{2^N-1},\\
\nu_{n>0}&=&e^{i\phi_n}\cos\theta_{2^N-1-n}\prod_{l=2^N-n}^{2^N-1}\sin\theta_l.
\end{eqnarray}
and 
\begin{equation}
\theta_n:=\arcsin\left(\xi_n^{\frac{1}{2n}}\right).
\end{equation}
The independent random variables $\phi_{n\geq 1}$ and $\xi_{n\geq 0}$ are uniformly distributed 
in the intervals:
\begin{equation*}
\phi_{n\geq 1}\in[0,2\pi), \qquad \xi_n\in[0,1].
\end{equation*}

\subsection{ Entangled states selection}

After generating a random  $N$-qubit state  $  |\psi\rangle$ we check  that it is actually  $N$-partite entangled. This happens iff for every bipartition $A/\hat{A}$ {(where $\hat{A}$ denotes the complement set of $A$)} of the $N$-qubit, ${\cal C}_G(\rho_A) \neq 0$, where  ${\cal C}_G$ is the generalized concurrence defined in \cite{GenConcurrence} as follows:
\begin{equation}
 \mathcal{C}_G(\rho_A):=2\sqrt{1-{\rm Tr}(\rho_A^2)}. 
\end{equation}

\subsection{Distances calculation}

It is possible to generate barcodes {{for simplicial complexes corresponding to a points (i.e.qubits) cloud by}}  giving in input {{to the persistent homology algorithm }} the  matrix $D$ of all pairwise distances between points. 
 \\
 In \cite{PH1}, a {{semi-distance $1/E_{i,j}, $}} was proposed,
 where $ E_{i,j} $ is an entanglement monotone calculated between qubit $ i $ and qubit $ j$. {{This semi-distance}}  goes from 1 (when the two qubit are maximally entangled) to +$\infty$ (when they are separable). 
{{Here}} we use the following semi-metric:
\begin{equation}
D_{i,j}=1-\exp\left\lbrace 1-\dfrac{1}{C_{i,j}}\right\rbrace, 
\end{equation}
where $ C_{i,j} $  is the concurrence between qubit $ i $ and qubit $ j $.  The semi-distance  $ D_{i,j} $  {{goes from $0$ to $1$ as the entanglement decreases, and remains finite for separable states}}.

Recall that, given a state $ \rho $ on $N$ qubits,  the concurrence  between two qubits $  i $ and $  j $  is obtained by first tracing out all other $ N-2 $ qubits. 
This gives the reduced density matrix  $ \rho _{ij}$.
Then 
\begin{equation}
C_{i,j}:= \max\{0,\lambda_1-\lambda_2-\lambda_3-\lambda_4\},
\end{equation}
where  $ \lambda_1,\lambda_2,\lambda_3,\lambda_4 $, are the square root of the eigenvalues 
(in decreasing order) of the matrix {{$ \rho_{ij}(\sigma_y\otimes\sigma_y)\rho^{*}_{ij}(\sigma_y\otimes\sigma_y)$ }} \cite{Concurrence}, {{with $\sigma_y$  the well known Pauli matrix and $\rho^{*}_{ij}$ the complex conjugate of $\rho_{i,j}$ in the computational basis.}}


\section{Entanglement classification}
\label{classification}

We have used the TDA package for computing persistent homology and barcodes developed  for the  R software. The classification is obtained grouping together  those states with the same barcode.

In the following barcodes, black lines represent connected components (i.e. homology group $ H_0 $),
red lines represent holes (i.e. homology group $ H_1 $) and blue lines represent voids (i.e. homology group $ H_2 $). All barcodes are generated using the Rips {{complex}}.

\subsection{Classification of  four qubits states}
Barcodes generated by 4-partite entangled states of 4 qubits and relative frequencies are shown starting  from the most frequent to the  least frequent one.
\begin{figure}[htbp]		
	\hspace{-1cm}
	\includegraphics[scale=0.4]{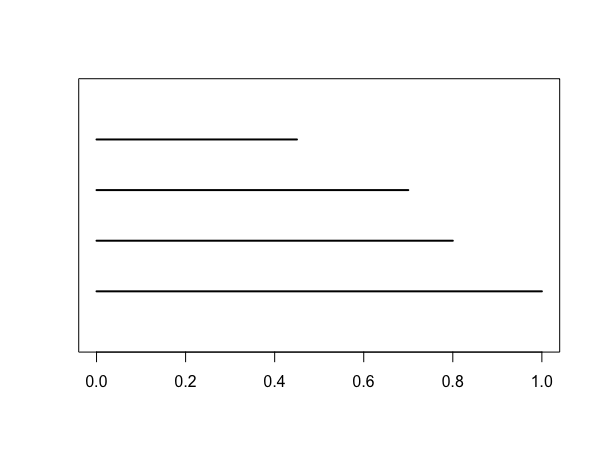}\vspace{-1cm}
	\caption{Barcode of   the class labelled as \textbf{4B1} }\label{fig:4B1}
\end{figure}
\FloatBarrier

Genuine entangled states with the   barcode shown in Figure~\ref{fig:4B1}  have a total of four connected components:  three of them    end at  value of $ \epsilon<1 $, while  only one  component persists over all the range of $ \epsilon $.
 The fact that only one connected component persists means that the state form a  single cluster  of qubits grouped by pairwise entanglement  without showing higher homological features.  A representative of such class {{is}} the $ |W\rangle  $ state.
{{
\begin{equation*}
|4, B1\rangle=|W\rangle=\frac{1}{2}(|0001\rangle+|0010\rangle+|0100\rangle+|1000\rangle)
\end{equation*}
}}
\begin{figure}[htbp]
	\hspace{-1cm}
	\includegraphics[scale=0.4]{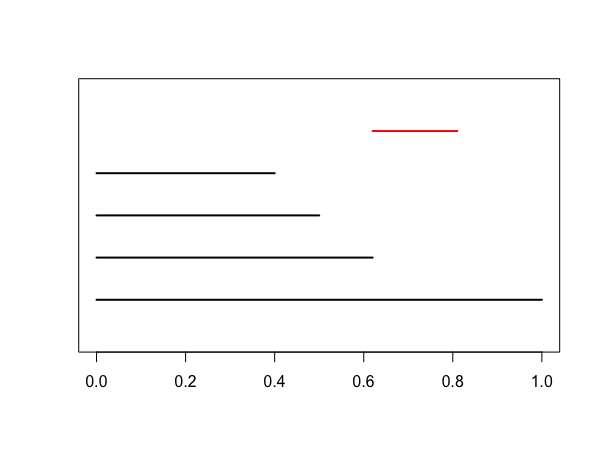}  \vspace{-1cm}
		\caption{Barcode of   the class labelled as  \textbf{4B2}}\label{fig:4B2}
\end{figure}\FloatBarrier
States belonging to  class of Figure~\ref{fig:4B2} form  again a single persistent component of pairwise entanglement between  qubits. However in this case a hole,  denoted by the red bar   $ H_1 $,  appears  when  only one connected  component is left. Such a  hole has limited life-span since   disappears when  $\epsilon$ is  sufficiently large. A state showing such a behaviour is the following:
{{
\begin{equation*}
\begin{split}
|4,B2\rangle=\dfrac{1}{2\sqrt{2}}\left( \sqrt{2}|0000\rangle+|0011\rangle+|0110\rangle+|1001\rangle+\right. \\ \left. +|1100\rangle+\sqrt{2}|1111\rangle  \right)
\end{split}
\end{equation*}
}}

\begin{figure}[htbp]
		\hspace{-1cm}
	\includegraphics[scale=0.4]{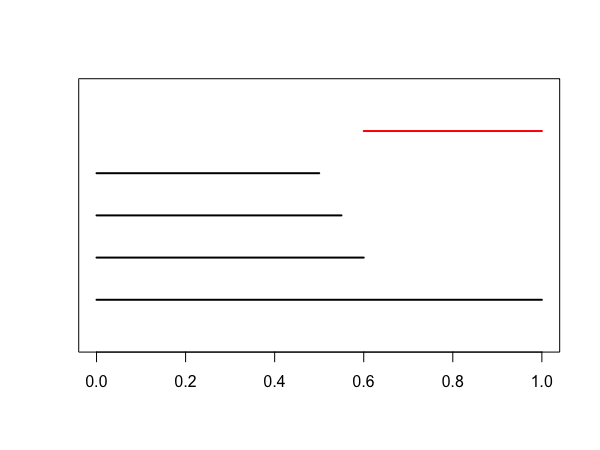}\vspace{-1cm}
		\caption{Barcode of   the class labelled as   \textbf{4B3}} \label{fig:4B3}
\end{figure}
\FloatBarrier
In the case shown in Figure~\ref{fig:4B3}, a single connected component is left and a persistent hole is present.
States with this barcode have the characteristic that  each qubit  is pairwise entangled   to other two qubits and completely   un-entangled  with a third qubit. A state showing   such properties is

\begin{equation*}
|4,B3\rangle={{\frac{1}{2}}}\left( |0000\rangle+|0011\rangle+|1010\rangle+|1111\rangle \right)
\end{equation*}

\begin{figure}[htbp]
	
		\hspace{-1cm}
	\includegraphics[scale=0.4]{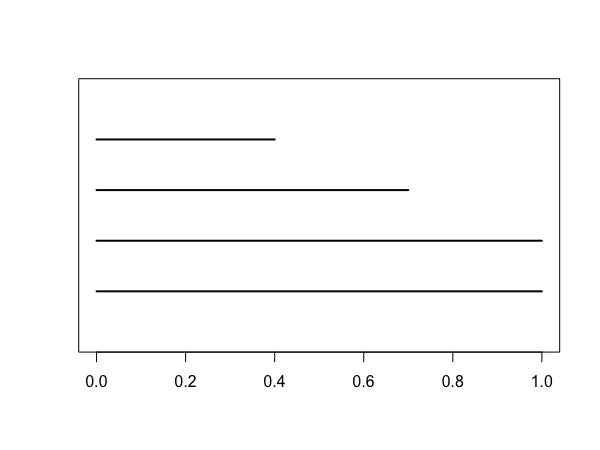}\vspace{-1cm}
		\caption{Barcode of   the class labelled as  \textbf{4B4}} \label{fig:4B4}
	\centering
\end{figure}
\FloatBarrier
In the class represented by the barcode in   Figure~\ref{fig:4B4} we find genuinely entangled states with no higher homological feature than $ H_0 $ which  have two different connected component that persist over the range of $\epsilon$. This means  that such states have two sets of qubits which are internally connected by pairwise entanglement to form a component, but no connection is present among qubits of different sets. Yet a single qubit in a set could be entangled to the other set as a whole. An example for this class is the state:
{{
\begin{equation*}
|4,B4\rangle=\frac{1}{2}\left(|0011\rangle+|1011\rangle+|1101\rangle+|1110\rangle\right)
\end{equation*}
  }}
\begin{figure}[htbp] 
		\hspace{-1cm}
\includegraphics[scale=0.4]{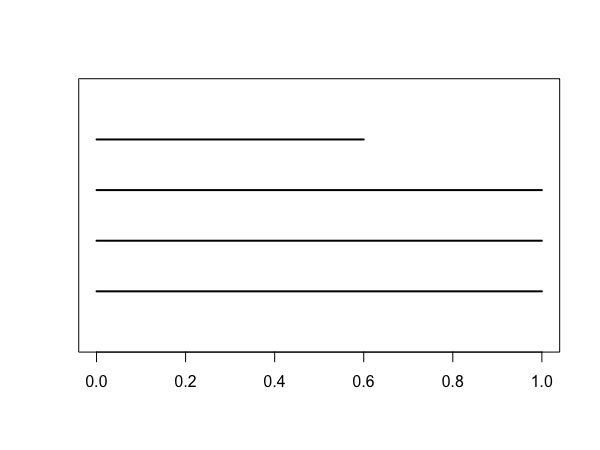}\vspace{-1cm}
		\caption{Barcode of   the class labelled as  \textbf{4B5}} \label{fig:4B5}
\end{figure}
\FloatBarrier
Like  the previous case, states with the barcode of Figure~\ref{fig:4B5} only show four  connected components, three of which persist  while one has  limited lifetime. 
The characteristic of these states is that there are always  2 qubits which do not share any pairwise entanglement with another qubit, while the other two do.
A representative state for this class is
{{
  \begin{equation*}
|4,B5\rangle=\dfrac{1}{\sqrt{3}}\left(|0000\rangle+|0111\rangle+|1101\rangle\right)
\end{equation*}
}}
\begin{figure}[htbp]
		\hspace{-1cm}
\includegraphics[scale=0.4]{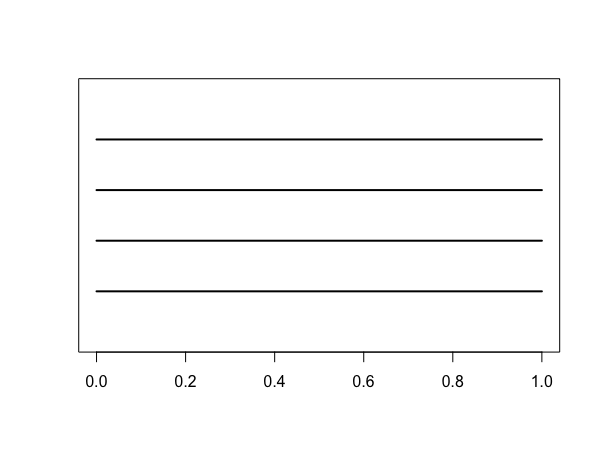}  \vspace{-1cm}
		\caption{Barcode of   the class labelled  as \textbf{4B6}} \label{fig:4B6}
\end{figure}
\FloatBarrier

\begin{figure}[htbp] 
	\centering
	\includegraphics[scale=0.6]{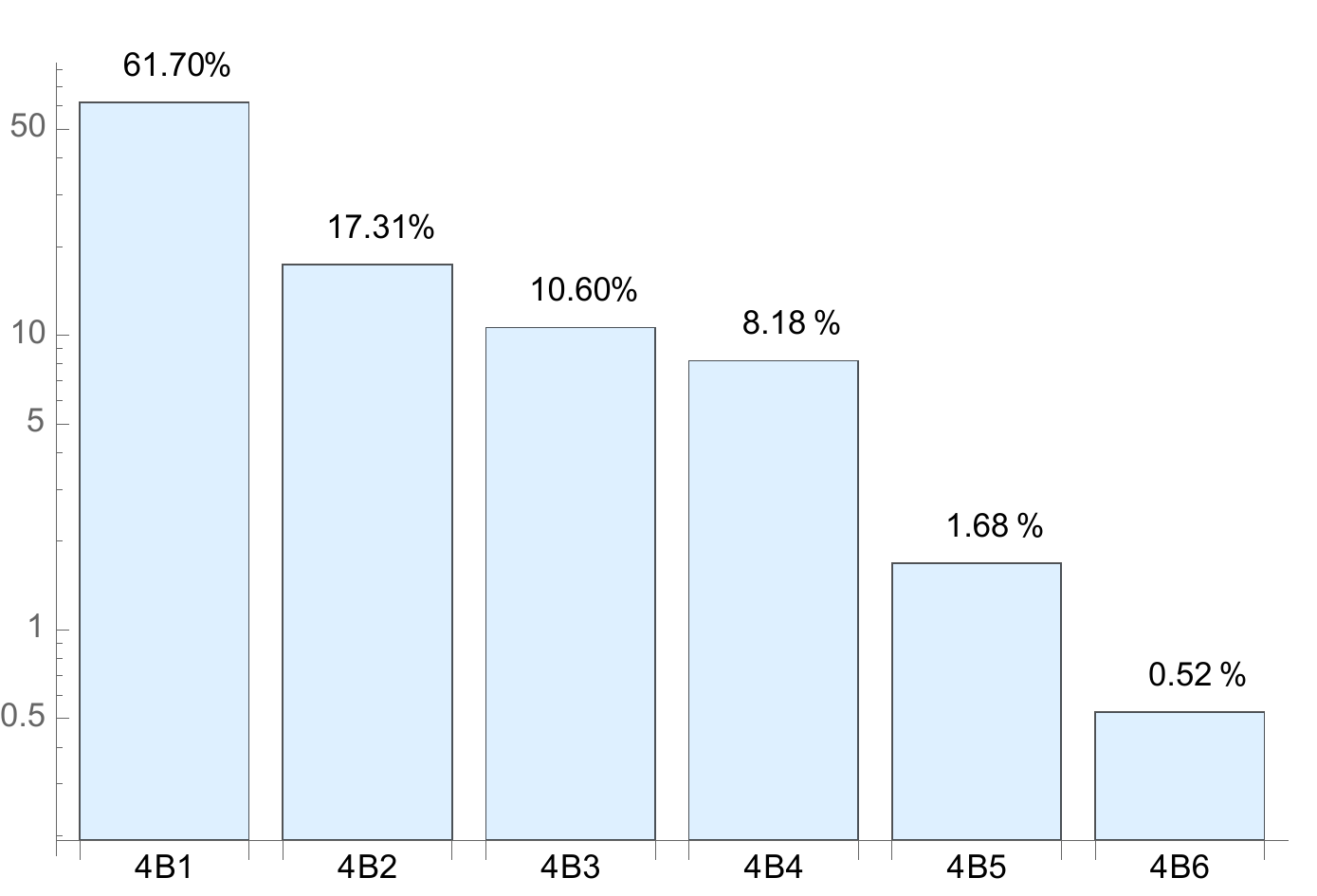}
	\caption{ Barcode frequencies (in Log scale) for four  qubits genuine entangled states   }
	\label{fig:4chart}
\end{figure}

States of the kind shown in Figure~\ref{fig:4B6}   do not have any pairwise entanglement among qubits. For this reason no qubit get connected to another and we see four distinct components that persist.
A representative of this class is the  {{$ |\text{GHZ}\rangle $}} state.
{{
\begin{equation*}
|4, B6\rangle=|\text{GHZ}\rangle=\frac{1}{\sqrt{2}}(|0000\rangle+|1111\rangle)
\end{equation*}
}}
{ As we can observe in Fig.\ref{fig:4chart},  there exist six different classes of four qubit genuine entangled states, based on the persistent homology classification.
The most frequent class ($ 61.70\% $) is the one where only one component persists (\textbf{4B1}), states like W belong to to this class.  With  frequencies of $ 17.31\% $ and $ 10.60\% $  we find states with barcodes \textbf{4B2} and \textbf{4B3} showing one persistent connected component and a hole (red line) that in the  case of  \textbf{4B3} is also persistent. The last three barcodes, in order  \textbf{4B4}, \textbf{4B5}  and \textbf{4B6} show an increasing number of disconnected components.  States that are GHZ-like are hence the least frequent ($ 0.52\% $).}
\FloatBarrier

\subsection{Classification of five  qubits states}

Let's now consider randomly generated  5-partite entangled states of 5 qubits. 
Barcodes and  relative frequencies are shown below starting  from the barcode more likely to appear to the    least frequent one.
\begin{figure}[htbp]
	\centering
	\hspace{-1cm}
	\includegraphics[scale=0.4]{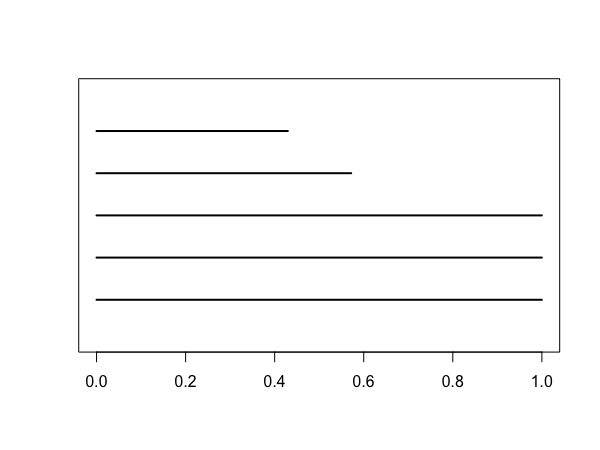} 	\vspace{-1cm}
	\caption{Barcode of   the class labelled  as \textbf{5B1}} \label{fig:5B1}
\end{figure}
\FloatBarrier
In the five qubit case, the most frequent class shows a barcode like the one in Figure~\ref{fig:5B1} with three  connected components that persist in the range of $ D $.
States of this kind have at least one qubit (up to  two qubits) that does not share pairwise entanglement with any  other qubit.  This configuration does not generate  higher homology groups~than~$H_0 $. An example of state in this class is 
  \begin{equation*}
  \begin{split}
|5,B1\rangle= \dfrac{1}{\sqrt{5}}\left( |00001\rangle+|00011\rangle+\right. \\ \left. + |00110\rangle+|01000\rangle+|11011\rangle\right)
  \end{split}
\end{equation*}
\begin{figure}[htbp]
		\hspace{-1cm}
	\includegraphics[scale=0.4]{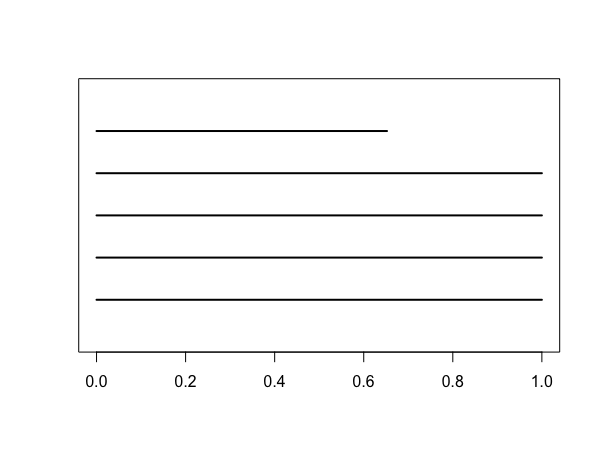} 	\vspace{-1cm}
	\caption{Barcode of   the class labelled  as \textbf{5B2}} \label{fig:5B2}
  \end{figure}
\FloatBarrier
As we can see,  the  barcode of Figure~\ref{fig:5B2} shows  four persistent connected components i.e. only two qubits among five  share pairwise entanglement while all remaining qubits act as independent connected component. A  representative  state for this  class is
  \begin{equation}
\begin{split}
|5,B2\rangle= \dfrac{1}{\sqrt{5}}\left( |00001\rangle+|00011\rangle+\right. \\ \left. + |00100\rangle+|01100\rangle+|11010\rangle\right) 
\end{split}
\end{equation}
\begin{figure}[htbp]
		\hspace{-1cm}
	\includegraphics[scale=0.4]{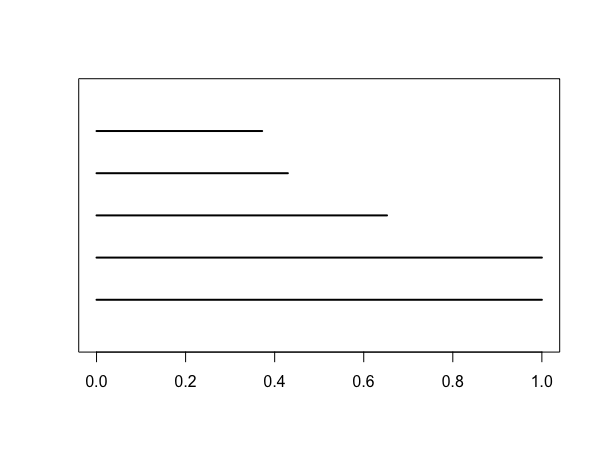} 	\vspace{-1cm}
	\caption{Barcode of   the class labelled  as \textbf{5B3}} \label{fig:5B3}
\end{figure}
\FloatBarrier
In this class identified by the barcode of Figure~\ref{fig:5B3}, two connected components persist while the other three have limited lifetime. {{Two}} clusters of qubits connected by pairwise entanglement are hence created and no holes or higher topological features appear.
An example of state in this class is the following
  \begin{equation*}
\begin{split}
|5,B3\rangle= \dfrac{1}{\sqrt{5}}\left( |00001\rangle+|00010\rangle+\right. \\ \left. + |00100\rangle+|01000\rangle+|10111\rangle\right)
\end{split}
\end{equation*}

\begin{figure}[htbp]
	\hspace{-1cm}
	\includegraphics[scale=0.4]{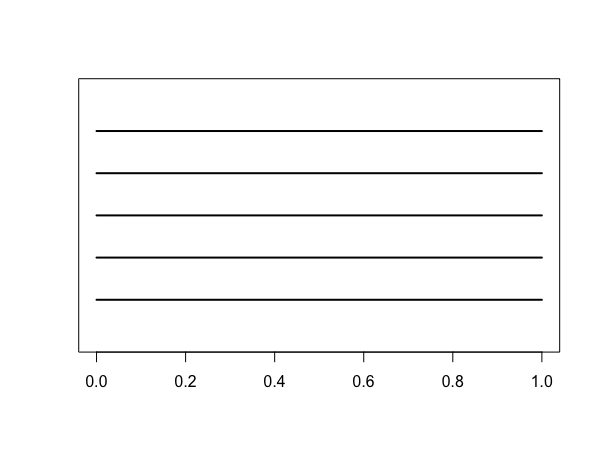} 	\vspace{-1cm}
	\caption{Barcode of   the class labelled  as \textbf{5B4}} \label{fig:5B4}
  \end{figure}
\FloatBarrier
Figure~\ref{fig:5B4}  show the barcode of the class where  we find GHZ like states, i.e. those  states where there are  no entangled pair of qubits and hence show five persistent connected components in the barcode. 
  \begin{equation*}
\begin{split}
|5,B4\rangle= \dfrac{1}{\sqrt{2}}\left( |00000\rangle+|11111\rangle\right)
\end{split}
\end{equation*}

\begin{figure}[htbp]
		\hspace{-1cm}
	\includegraphics[scale=0.4]{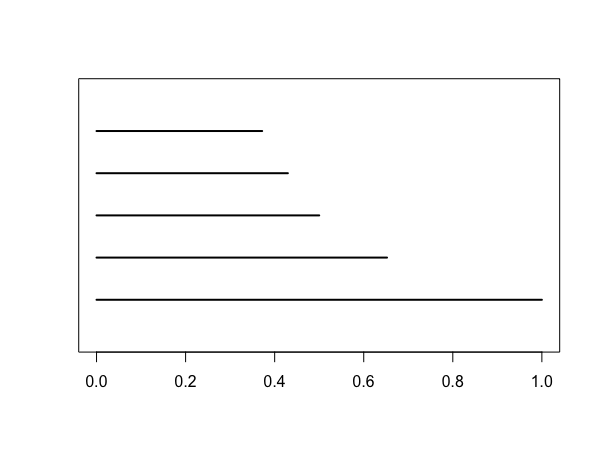} 	\vspace{-1cm}
		\caption{Barcode of   the class labelled  as \textbf{5B5}} \label{fig:5B5}
		\centering
\end{figure}
\FloatBarrier
After we find the barcode shown in Figure~\ref{fig:5B5} and   relative to those states like $ |W\rangle $ which have only one connected component and hence pairwise entanglement creates a single cluster of qubits. 
  \begin{equation*}
\begin{split}
|5,B5\rangle= \dfrac{1}{\sqrt{5}}\left( |00001\rangle+|00010\rangle+\right. \\ \left. + |00100\rangle+|01000\rangle+|10000\rangle\right)
\end{split}
\end{equation*}

	\begin{figure}[htbp]
			\hspace{-1cm}
	\includegraphics[scale=0.4]{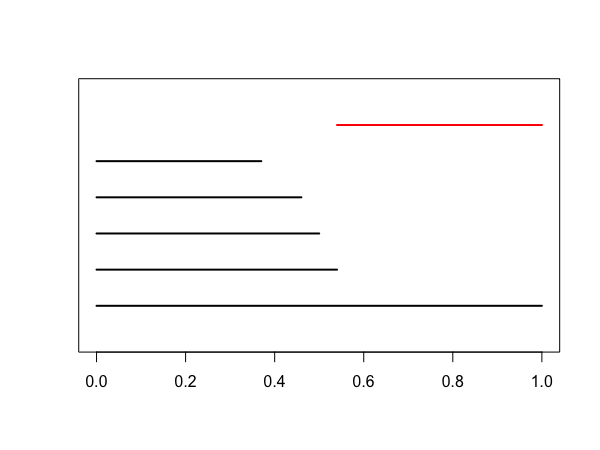}  	\vspace{-1cm}
		\caption{Barcode of   the class labelled  as \textbf{5B6}} \label{fig:5B6}
\end{figure}
\FloatBarrier
Figure~\ref{fig:5B6}  shows the barcode of the first class of 5 qubits genuinely entangled states that present a first order  homology group $ H_1 $, i.e. a hole, in the barcode.
States in this class have their qubits connected to form a single persistent component when $ \epsilon \approx 1 $. Note also that some    subsets of qubits  do not share any pairwise entanglement  and hence are responsible for the persistent hole.
An example of state in this class is 
  \begin{equation*}
\begin{split}
|5,B6\rangle= \dfrac{1}{\sqrt{6}}\left( |00000\rangle+|11000\rangle+\right. \\ \left. + |01100\rangle+|00110\rangle+|00011\rangle+|10001\rangle \right)
\end{split}
\end{equation*}

\begin{figure}[htbp]
	\hspace{-1cm}
	\includegraphics[scale=0.4]{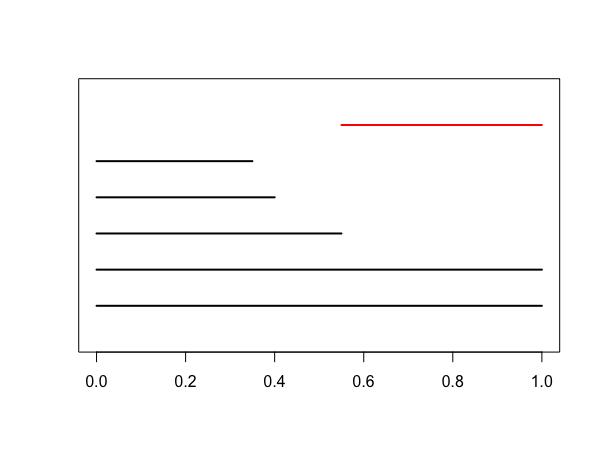} 	\vspace{-1cm}
		\caption{Barcode of   the class labelled  as \textbf{5B7}} \label{fig:5B7}
\end{figure}
\FloatBarrier
As we can see in Figure~\ref{fig:5B7}, like in the previous class, a hole is created at some value of $\epsilon$ and persists up to the upper limit of the semi-metric $D$. However here while four qubits are responsible for one connected component and  for the $ H_1 $ homology,  the remaining fifth  qubit  does not share any pairwise entanglement with the others and creates a persistent component on its own.
A representative state of this kind is 
\begin{equation*}
\begin{split}
|5,B7\rangle=\dfrac{1}{\sqrt{5}}\left( |00010\rangle+|00011\rangle+\right. \\ \left. + |00101\rangle+|10111\rangle+|11011\rangle \right)
\end{split}
\end{equation*}

\begin{figure}[htbp]
	\hspace{-1cm}
	\includegraphics[scale=0.4]{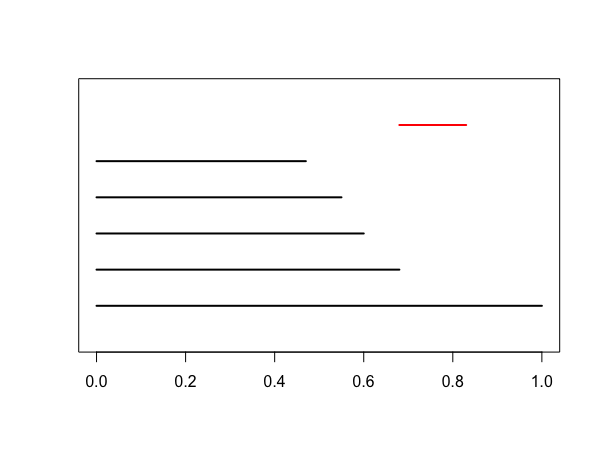} 	\vspace{-1cm}
		\caption{Barcode of   the class labelled  as \textbf{5B8}} \label{fig:5B8}
\end{figure}
\FloatBarrier
States in the class of Fig.\ref{fig:5B8} have similar properties to those in the class of Fig.\ref{fig:5B6}, however in this case the $ H_1 $ homology does not persist since connections among  qubits creating the hole appear   at some value of $\epsilon$. An example state with such barcode could be
  \begin{equation}
\begin{split}
|5,B8\rangle= \dfrac{1}{\sqrt{10}}\left( \sqrt{5}|00000\rangle+|11000\rangle+\right. \\ \left. + |01100\rangle+|00110\rangle+|00011\rangle+|10001\rangle \right). 
\end{split}
\end{equation}

\begin{figure}[htbp]
	\hspace{-1cm}
	\includegraphics[scale=0.4]{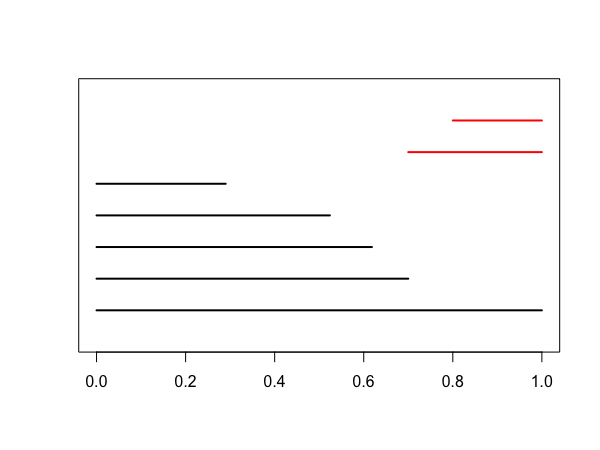} 	\vspace{-1cm}
		\caption{Barcode of   the class labelled  as \textbf{5B9}} \label{fig:5B9}
\end{figure}
\FloatBarrier
The class characterized by the barcode depicted in Figure~\ref{fig:5B9} shows a single persistent connected component of qubits grouped by pairwise entanglement but  also two holes  which  appear at some    $\epsilon$ and persist for higher values.
A representative for this class is the following
\begin{equation*}
\begin{split}
|5,B9\rangle= \sqrt{\dfrac{2}{5} }\ \left( |00000\rangle+|01010\rangle\right)  +\dfrac{1}{5}\left( |00011\rangle+\right. \\ \left. + |00101\rangle+|01100\rangle+|11000\rangle+|10001\rangle \right)
\end{split}
\end{equation*}

\begin{figure}[htbp]
	\hspace{-1cm}
	\includegraphics[scale=0.4]{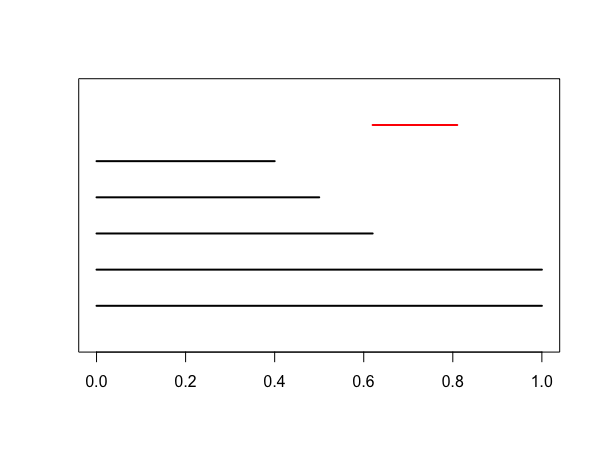}  	\vspace{-1cm}
		\caption{Barcode of   the class labelled  as \textbf{5B10}} \label{fig:5B10}
			\centering
\end{figure}
\FloatBarrier
States belonging to the    class  of Figure~\ref{fig:5B10} have similar properties to those in class with barcode in Figure~\ref{fig:5B7}, i.e. two persistent connected components, one of which is made up of a single qubit which does not share  pairwise entanglement with the other four. In the other  instead, the remaining four qubits   get  connected to form a  non-persistent  hole. An example state with such barcode could be
\begin{equation*}
\begin{split}
|5,B10\rangle= \dfrac{1}{\sqrt{6}}\left( |01000\rangle+|01010\rangle+\right. \\ \left. + |10000\rangle+|10001\rangle+|10110\rangle+|11010\rangle \right) 
\end{split}
\end{equation*}

	\begin{figure}[htbp]
	\hspace{-1cm}
	\includegraphics[scale=0.4]{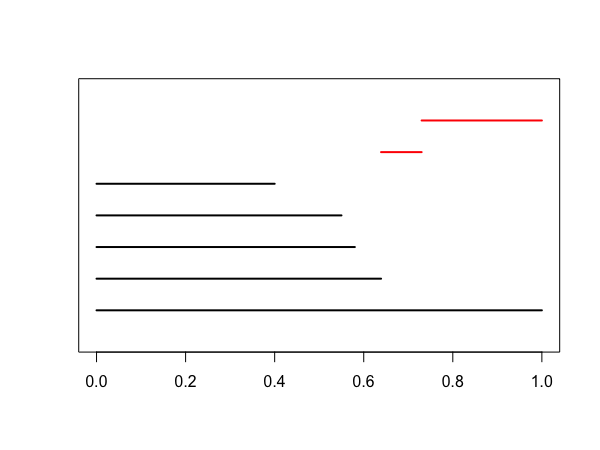} 	\vspace{-1cm}
	\caption{Barcode of   the class labelled  as \textbf{5B11}} \label{fig:5B11}
\end{figure}
\FloatBarrier
A single persistent connected component and two holes characterize the barcode of this class, as shown in Figure~\ref{fig:5B11}. Note that one of the two homology group generators $ H_1 $ appears only in a limited interval while the other one persists over $\epsilon$.
An example state with such barcode:
\begin{equation*}
\begin{split}
|5,B11\rangle= \dfrac{1}{\sqrt{11}}\left( \sqrt{5}|00010\rangle+\sqrt{2}|00100\rangle+\right. \\ \left. + \sqrt{2}|10000\rangle+|10101\rangle+|11100\rangle \right)
\end{split}
\end{equation*}

\begin{figure}[htbp]
	\hspace{-1cm}
	\includegraphics[scale=0.4]{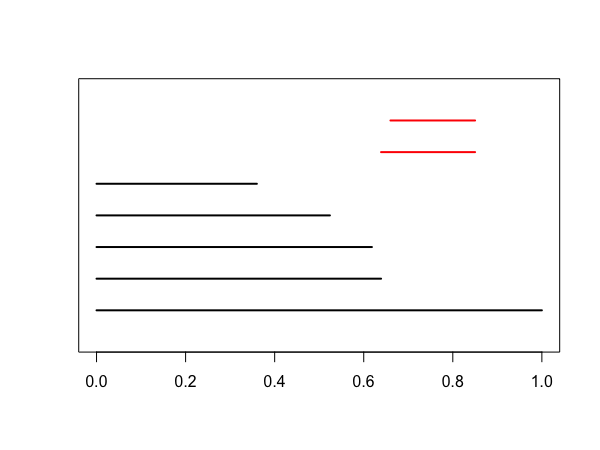} 	\vspace{-1cm}
		\caption{Barcode of   the class labelled  as \textbf{5B12}} \label{fig:5B12}
	\centering
\end{figure}
\begin{figure}[htbp]
	\centering
	\includegraphics[scale=0.65]{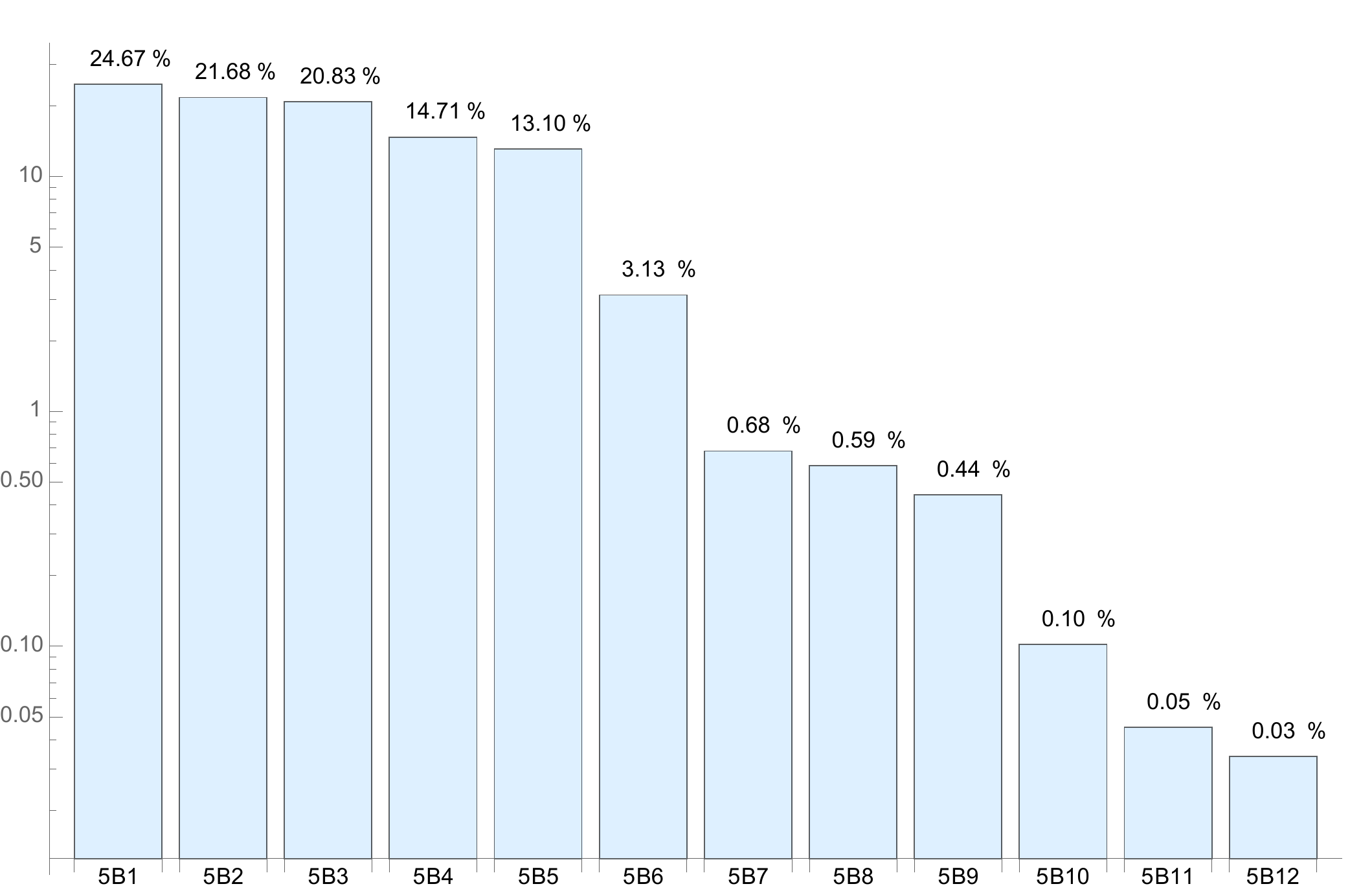}
\caption{ Barcode frequencies (in Log scale) for five qubits genuine entangled states   }
\label{fig:5chart}
\end{figure}

\FloatBarrier
The least frequent class, barcode in Figure~\ref{fig:5B12}, is the one composed of those states where qubits get connected to form a single  connected component  allowing the presence of two hole that however do not persist.
An example state with such barcode is the following
  \begin{equation*}
\begin{split}
|5,B12\rangle= \dfrac{1}{\sqrt{2} }\ |00000\rangle +\dfrac{1}{\sqrt{10}}\left( |11000\rangle+\right. \\ \left. + |01100\rangle+|01010\rangle+|00101\rangle+|10001\rangle \right)
\end{split}
\end{equation*}
{
By looking at the chart in Figure~\ref{fig:5chart} it can be  noticed that the most frequent barcode (\textbf{5B1}) belongs to states with three connected components, followed by states with barcodes showing four, two, five and one persistent components (respectively \textbf{5B2}, \textbf{5B3}, \textbf{5B4}, \textbf{5B5}).  The $ 95\% $ of all randomly generated  states fall inside one of these first five classes. After them, barcodes with higher dimensional homology  features start to appear: at first those with one hole and then those with two. The only exception is given by barcode  \textbf{5B10}  (showing one short-lived hole but two connected components) since it is less frequent than \textbf{5B9} (one connected component and two persistent holes).}

\onecolumngrid

\FloatBarrier

\twocolumngrid

\subsection{Classification of six qubits states}
\begin{figure}[htbp] \vspace{0cm}
	\includegraphics[scale=0.66]{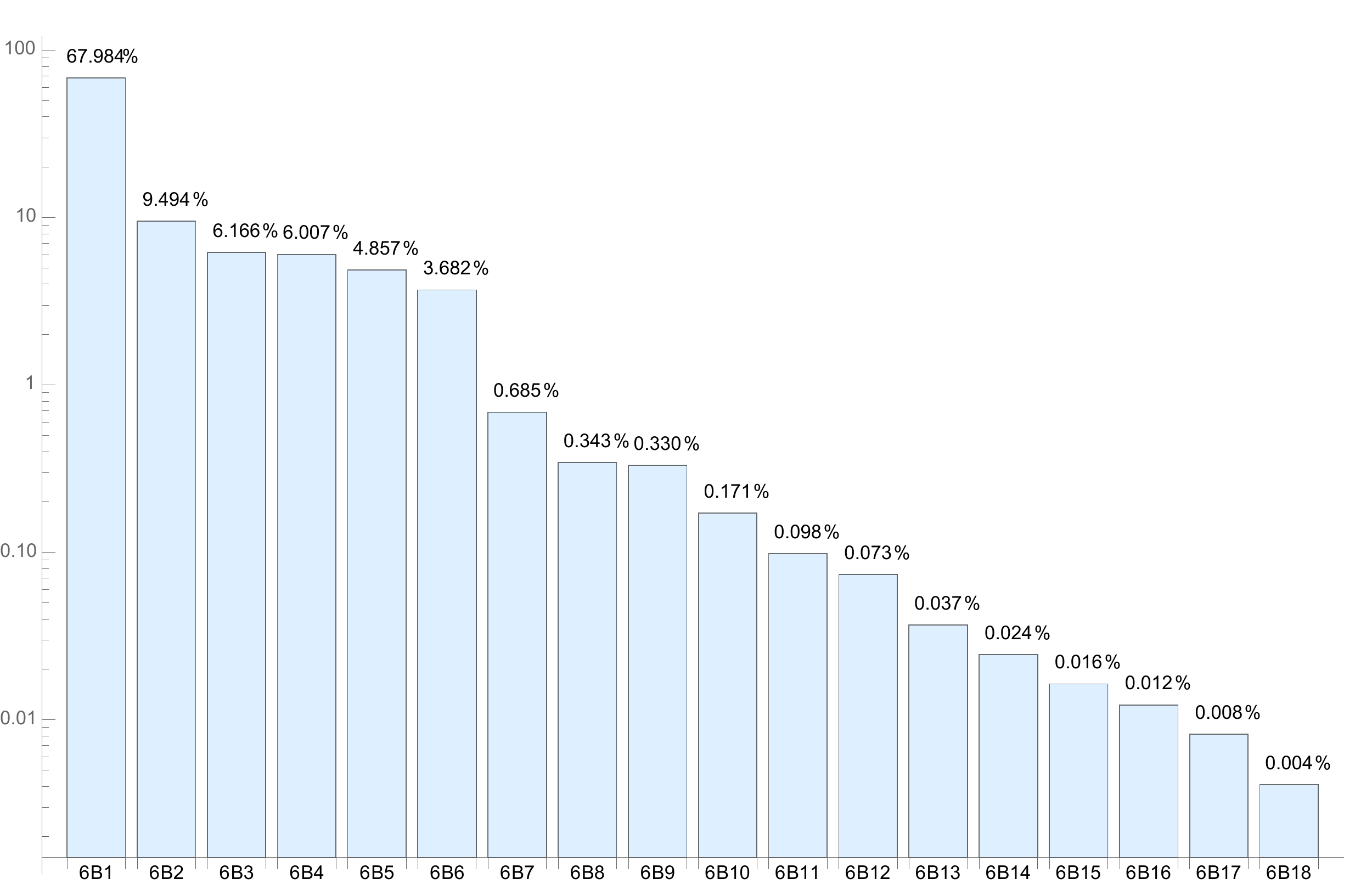}
	\caption{ Barcode frequencies (in Log Scale) for six qubits genuine entangled states   }
	\label{fig:6chart}
\end{figure}

 By randomly generating six-partite genuine entangled states of six qubits, we obtained 33 different classes. Their barcodes are presented in  Appendix \ref{appendixA}, while frequencies are shown in Figure~\ref{fig:6chart}. 

With a frequency of $ 68\% $, the class defined by barcode \textbf{6B1} is by far the most frequent. Such a class consists of GHZ-like  states that present  six different connected components, i.e.  the single qubits with no pairwise entanglement. 

As we have seen, for the five qubit case, the first classes in frequency, from  \textbf{6B1} to \textbf{6B6,} are those  containing states showing only connected components ($ H_0  $ homology group i.e. black bars). 

Then states with one hole start to appear, and later those showing  two holes, with the exception of barcodes \textbf{6B12} and \textbf{6B13}.

Finally,  barcodes showing multiple  holes and voids, from  \textbf{6B19} to \textbf{6B33} are also possible but they  do not appear in the histogram since their frequency is very low  ($\ll 0.004 \% $).
\\\\

\onecolumngrid

\FloatBarrier

\twocolumngrid 

\section{conclusion }
\label{conclusion}
The classification that we have carried out for four, five  and six qubits  entangled  states shows that is  possible to distinguish respectively  six , twelve and thirty-three different classes  by persistent homological barcodes.
In general, given a $ N $ qubit genuinely entangled state, it is always possible to come up with a finite classification where the total number of possible  barcodes $ B_N $ is bounded by
\begin{equation*}
B_N<\left( \sum_{e=0}^{\frac{N(N-1)}{2}}G_N(d)\ d!\right) 
\end{equation*}
where   $ G_N(d) $ is equal to the   number of all possible graphs with $ N $ vertices and $ d $ edges. The factorial $ d! $ is necessary to take  into consideration all  possible ways of building  $ G_N(d) $.
\\

Furthermore,  patterns  seems to emerge in our    classifications by looking at the   frequencies of barcodes. First of all,  those  states which are characterized by the only  $ H_0 $ homology group become more likely as the number of qubits $ N $ increases.  This  is followed by the group of those states showing also  $ H_1$ which again are followed by those with a much richer topology.
While this fact could be  explained from a topological point of view  by claiming that, with a limited number of points complex homological patterns in the barcode  are harder to obtain, it is still interesting to notice that the same reasoning also hold true for quantum state barcodes.
\\

Among those  states with only the   $ H_0 $ homology group, it is worth noticing that the  W-like class,  with only one persistent connected component, decreases its frequency with the increase of $ N $. In fact, except for  $ N=4 $ where  we find this class in the first place, in the $ N=5  $ and $ N=6 $ cases, it falls to the  last position.
Conversely, the class of states which have $ N $ persistent  components, like GHZ, gradually increase their frequency, starting from the bottom at $ N=4  $ and becoming the most popular at $ N=6 $.

In general we can say  that increasing the number of qubits makes the randomly generated states  easily fall inside  classes   with  more persistent  connected components. This seems to indicate  that, increasing the number of parties, qubits in a genuine entangled states tend to dislike pairwise entanglement and rather share it with the whole set  of other qubits.

One last consideration is related to the study of quantum algorithms and their complexity classes. As quantum speed-up is essentially based on the entanglement employed in the algorithm,  it would be subject of future work the study of the relation between quantum complexity  classes   and  the  persistent homology classification presented here.

\appendix
\section{Barcodes for six-partite genuine entangled states}
\label{appendixA}

Here we report the 33 different barcodes obtained from randomly generated six-partite genuine entangled states.  
As usual,  barcodes are presented  starting  from the most frequent to the  least frequent one. 

\begin{figure}[htbp]	
	\includegraphics[scale=0.3]{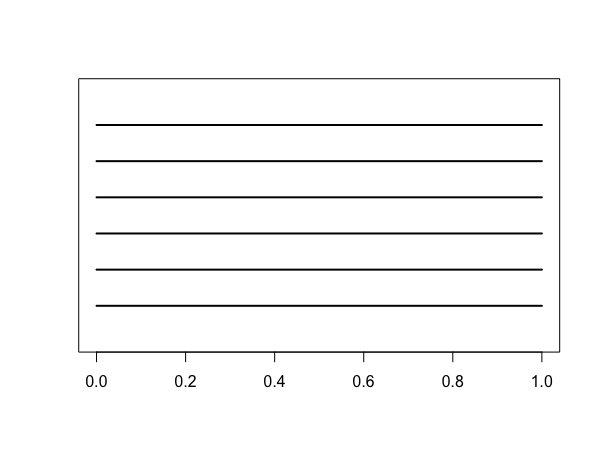} \hspace{-1cm}\\ \vspace{-0.5cm}
{\textbf{6B1}}
\\
	\includegraphics[scale=0.3]{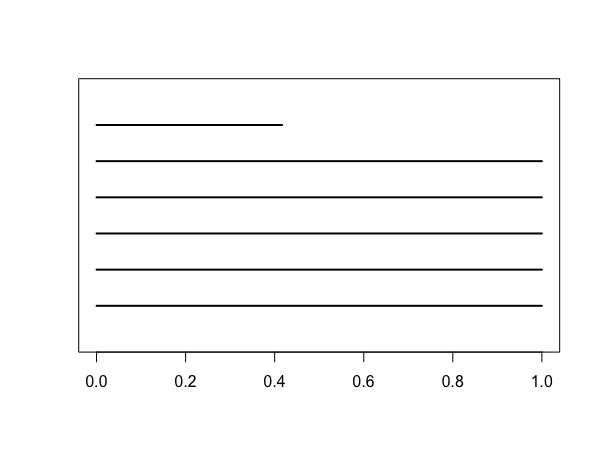}\hspace{-1cm}\\ \vspace{-0.5cm}
{\textbf{6B2}}
\\

	\includegraphics[scale=0.3]{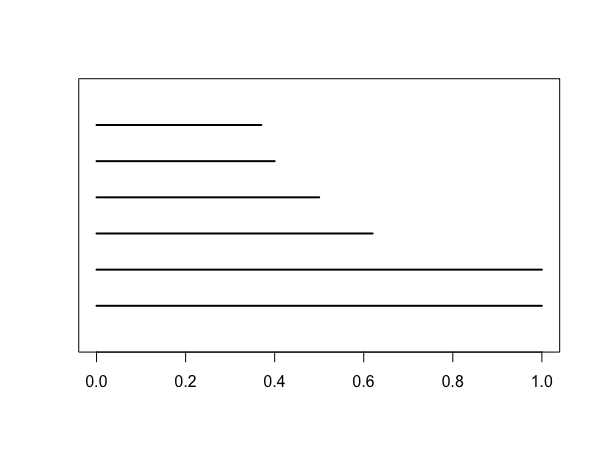}\hspace{-1cm}	\\ \vspace{-.5cm}
{\textbf{6B3}}
\\
	\includegraphics[scale=0.3]{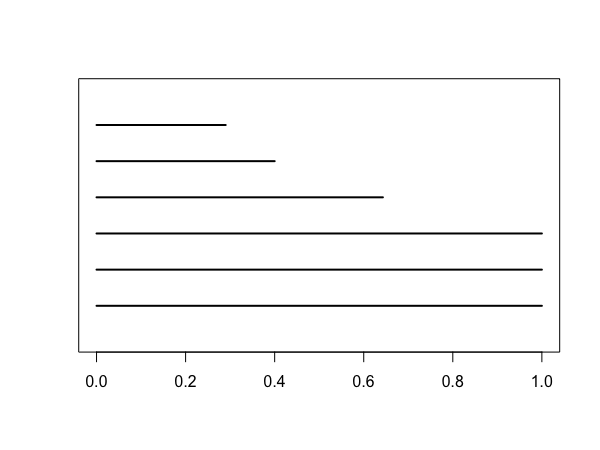}\hspace{-1cm}\\ \vspace{-.5cm}
{\textbf{6B4}}
\end{figure}

\begin{figure}[htbp]	
	\includegraphics[scale=0.3]{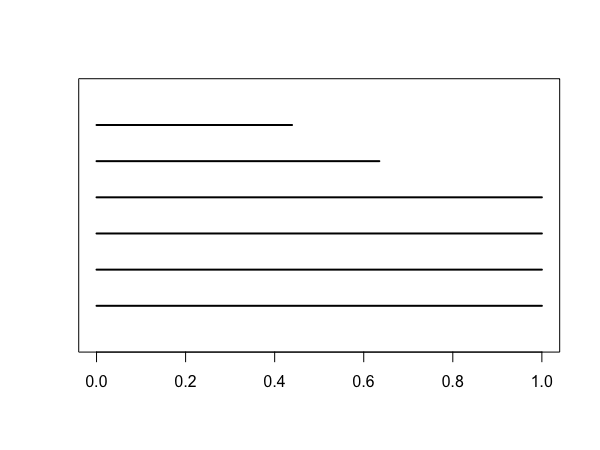}\hspace{-1cm}\\ \vspace{-.5cm}
{\textbf{6B5}}
\\
	\includegraphics[scale=0.3]{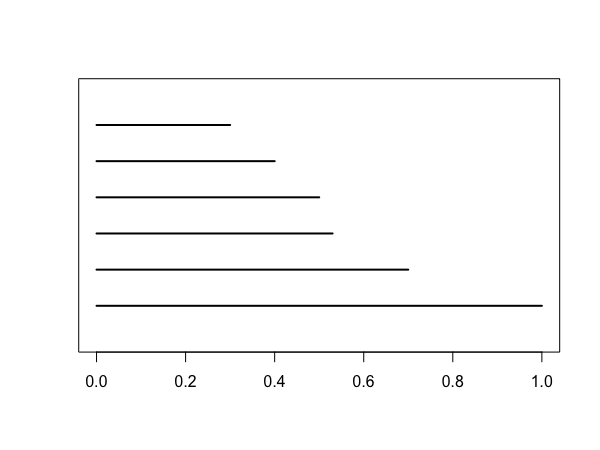}\hspace{-1cm}\\
\vspace{-.5cm}{\textbf{6B6}}
\\
	\includegraphics[scale=0.3]{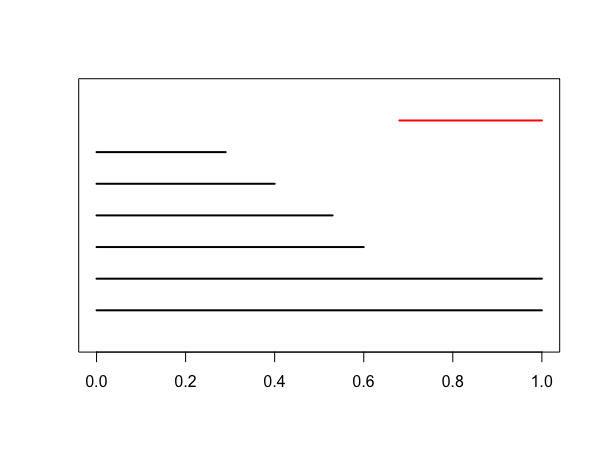}\hspace{-1cm}\\
	\vspace{-.5cm} {\textbf{6B7}}
\\
	\includegraphics[scale=0.3]{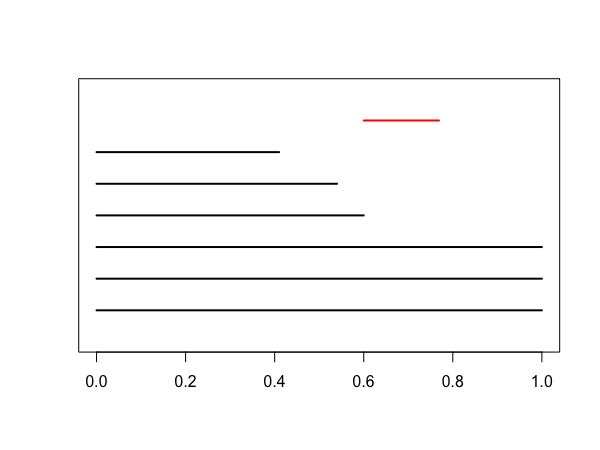}\hspace{-1cm}\\
	\vspace{-.5cm}{\textbf{6B8}}
\\
	\includegraphics[scale=0.3]{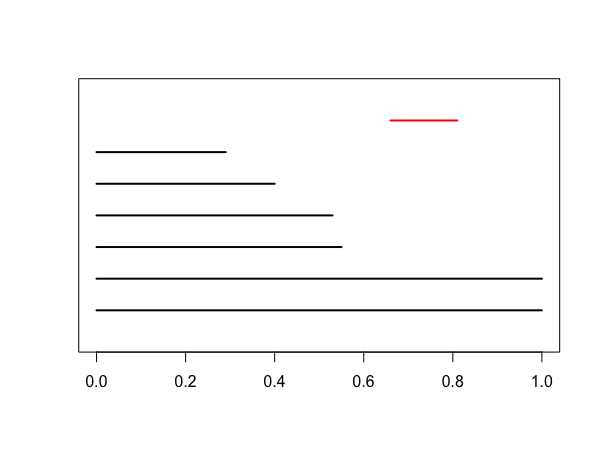}\hspace{-1cm}\\
	\vspace{-.5cm}{\textbf{6B9}}
\end{figure}

\begin{figure}[htbp]	
	\includegraphics[scale=0.3]{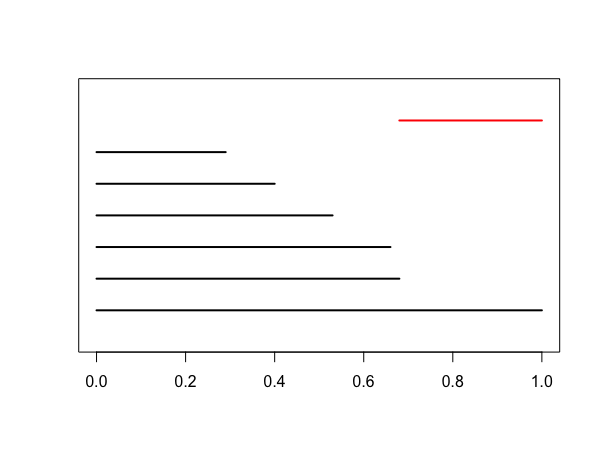}\hspace{-1cm}\\
	\vspace{-.5cm}{\textbf{6B10}}
	\\
		\includegraphics[scale=0.3]{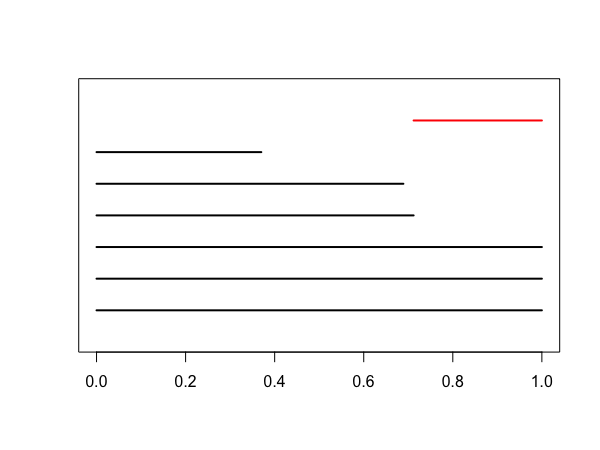}\hspace{-1cm}\\
	\vspace{-.5cm}{\textbf{6B11}}
	\\

	\includegraphics[scale=0.3]{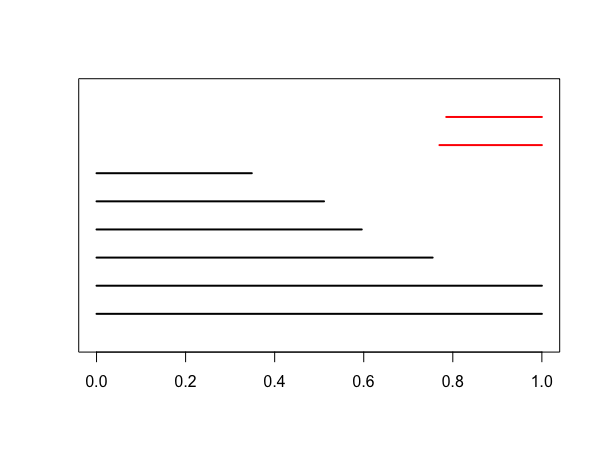}\hspace{-1cm}\\
	\vspace{-.5cm}{\textbf{6B12}}
\\
	\includegraphics[scale=0.3]{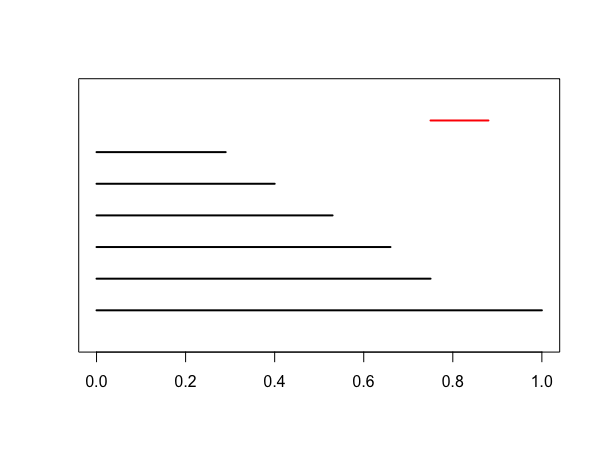}  \hspace{-1cm}\\
	\vspace{-.5cm}{\textbf{6B13}}
\\
	\includegraphics[scale=0.3]{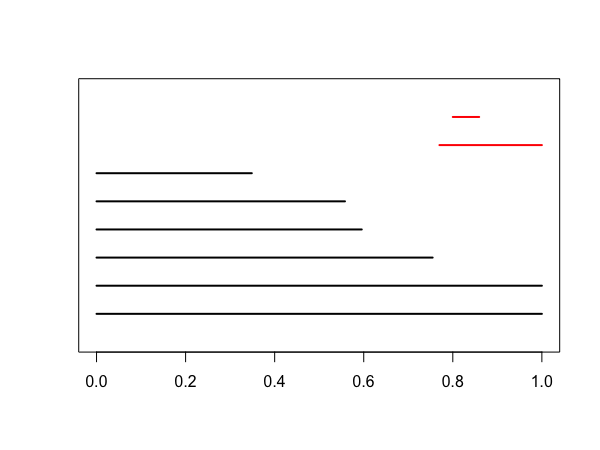}\hspace{-1cm}\\
	\vspace{-.5cm}{\textbf{6B14}}
\end{figure}

\begin{figure}[htbp]	
	\includegraphics[scale=0.3]{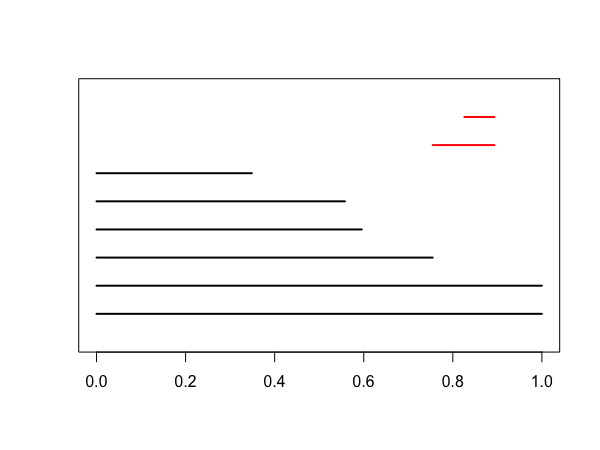}\hspace{-1cm}\\
	\vspace{-.5cm}{\textbf{6B15}}
	\\	
	
	\includegraphics[scale=0.3]{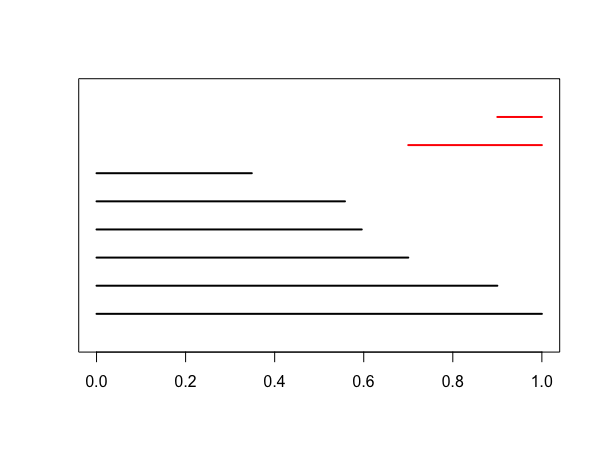} \hspace{-1cm}\\
	\vspace{-.5cm}{\textbf{6B16}}
	\\

	\includegraphics[scale=0.3]{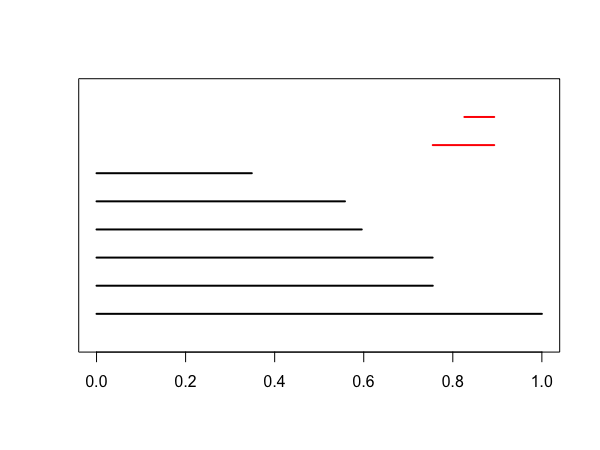}\hspace{-1cm}\\
	\vspace{-.5cm}{\textbf{6B17}}
\\
	\includegraphics[scale=0.3]{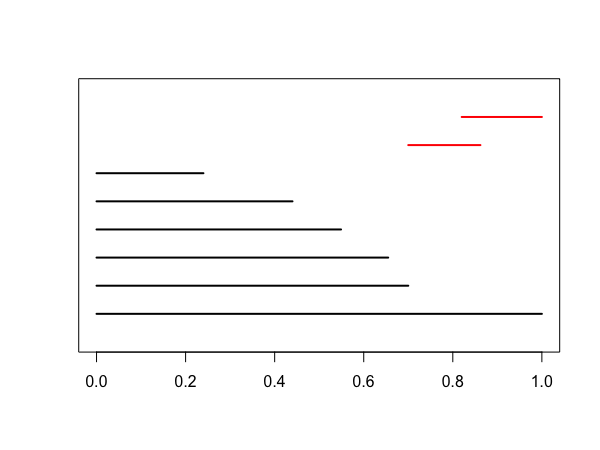} \hspace{-1cm}\\
	\vspace{-.5cm}{\textbf{6B18}}
\\
	\includegraphics[scale=0.3]{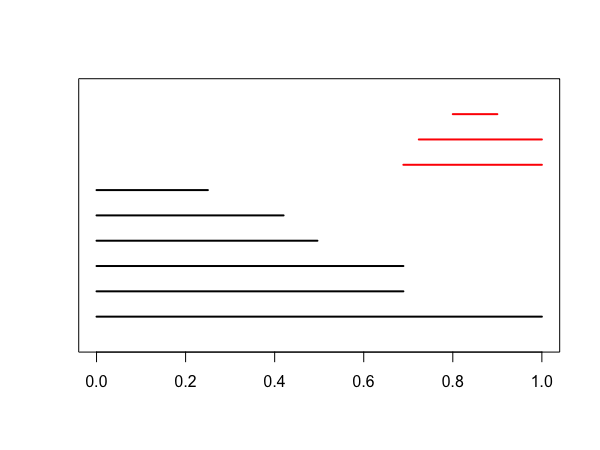}\hspace{-1cm}\\
	\vspace{-.5cm}{\textbf{6B19}}
\end{figure}

\begin{figure}[htbp]	
	\includegraphics[scale=0.3]{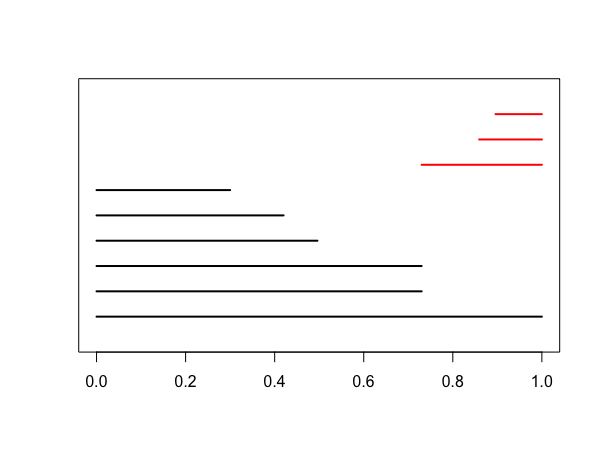}  \hspace{-1cm}\\
	\vspace{-.5cm}{\textbf{6B20}}
	\\
		\includegraphics[scale=0.3]{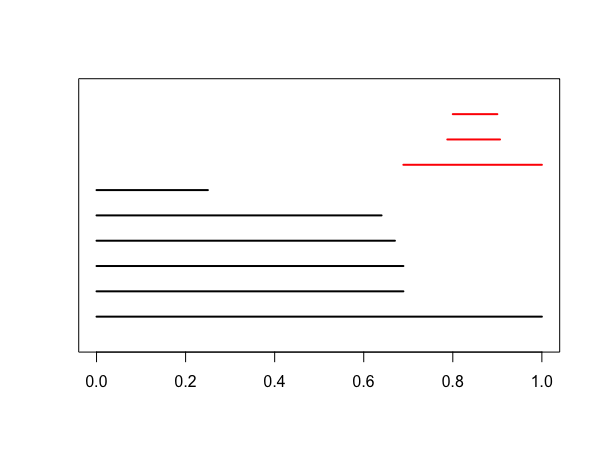}\hspace{-1cm}\\
	\vspace{-.5cm}{\textbf{6B21}}
	\\

	\includegraphics[scale=0.3]{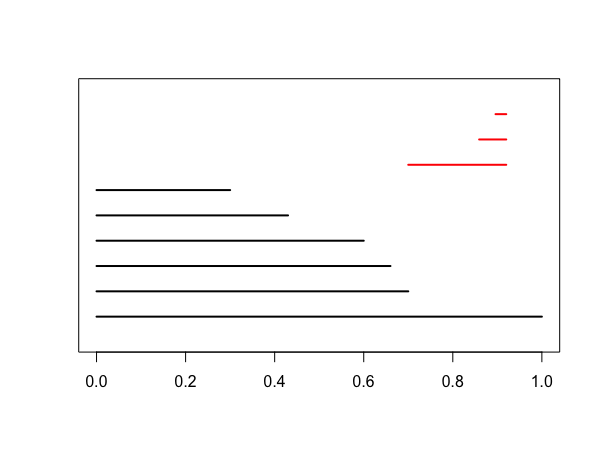}  \hspace{-1cm}\\
	\vspace{-.5cm}{\textbf{6B22}}
\\
	\includegraphics[scale=0.3]{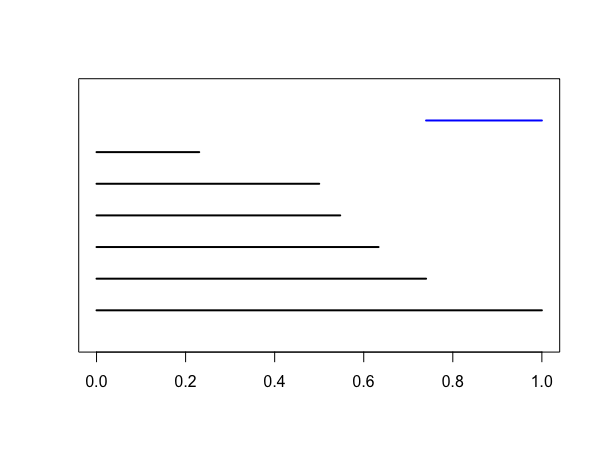} \hspace{-1cm}\\
	\vspace{-.5cm}{\textbf{6B23}}
\\
	\includegraphics[scale=0.3]{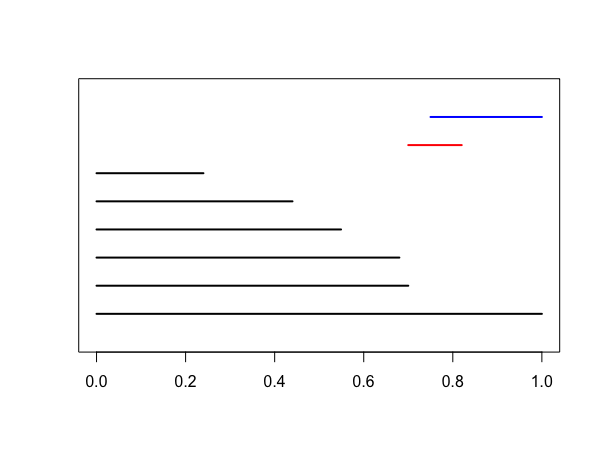} \hspace{-1cm}\\
	\vspace{-.5cm}{\textbf{6B24}}
\end{figure}

\begin{figure}[htbp]	
	\includegraphics[scale=0.3]{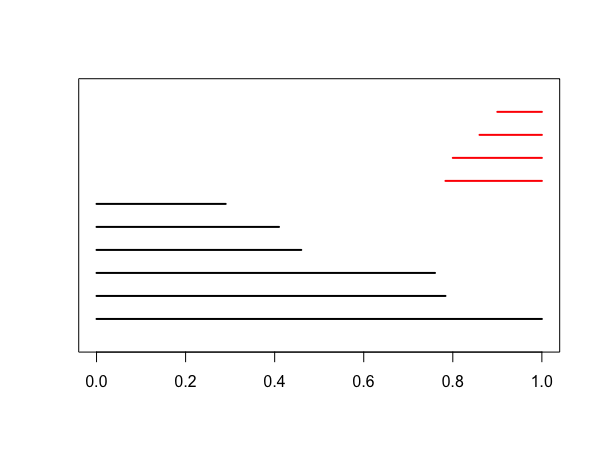} \hspace{-1cm}\\
	\vspace{-.5cm}{\textbf{6B25}}
	\\
		\includegraphics[scale=0.3]{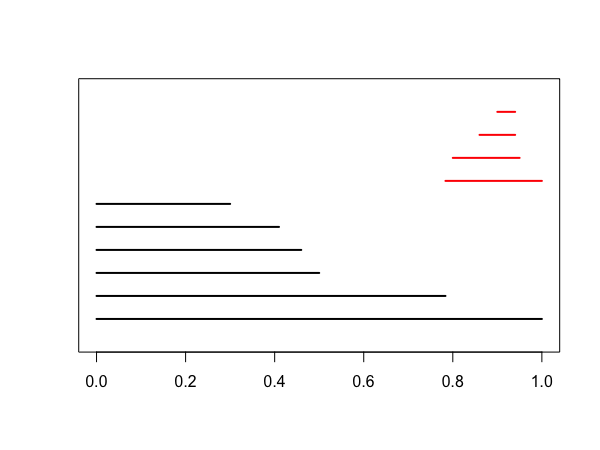} \hspace{-1cm}\\
	\vspace{-.5cm}{\textbf{6B26}}
	\\

	\includegraphics[scale=0.3]{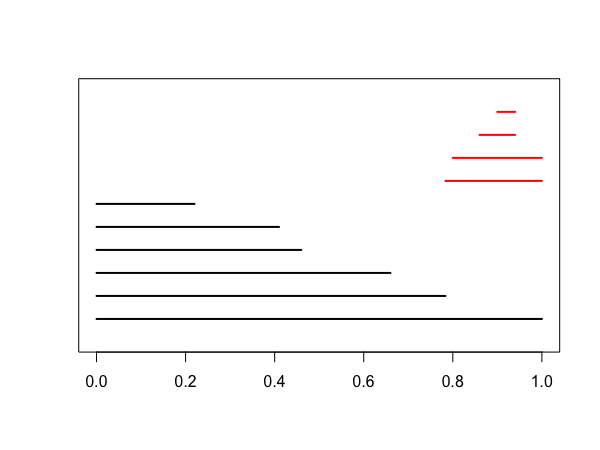} \hspace{-1cm}\\
	\vspace{-.5cm}{\textbf{6B27}}
\\
	\includegraphics[scale=0.3]{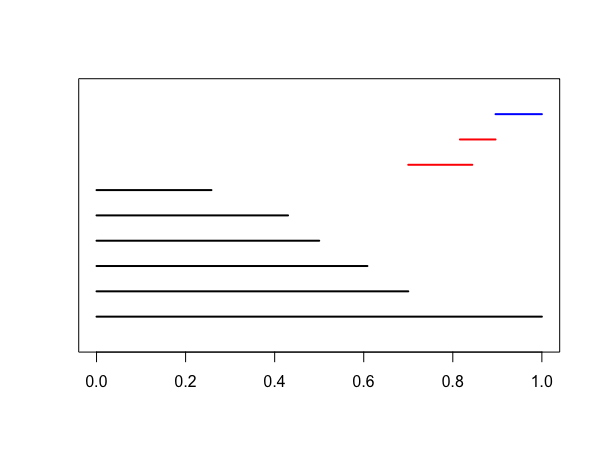} \hspace{-1cm}\\
	\vspace{-.5cm}{\textbf{6B28}}
\\
	\includegraphics[scale=0.3]{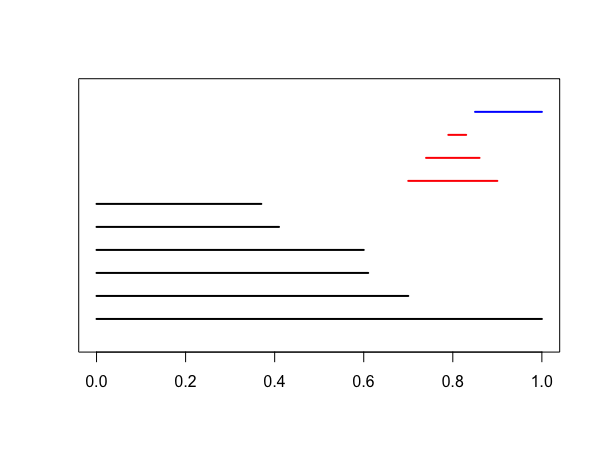} \hspace{-1cm}\\
	\vspace{-.5cm}{\textbf{6B29}}
\end{figure}

\begin{figure}[htbp]	
	\includegraphics[scale=0.3]{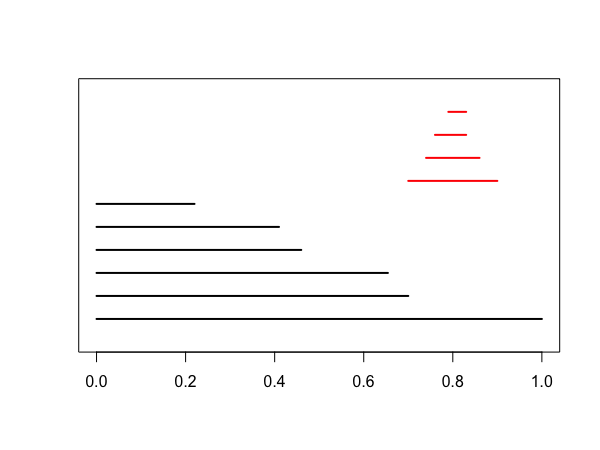} \hspace{-1cm}\\
	\vspace{-.5cm}{\textbf{6B30}}
\\
	\includegraphics[scale=0.3]{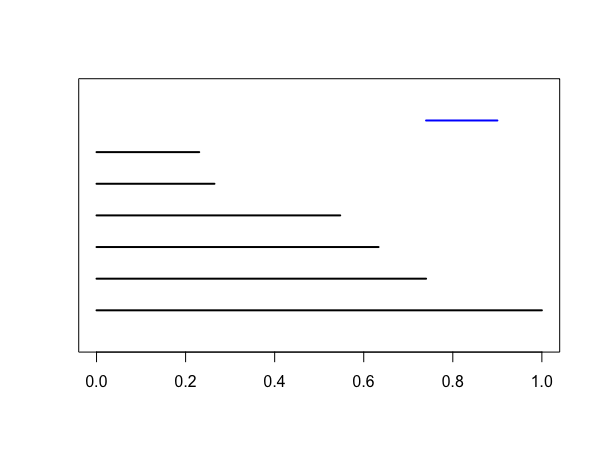} \hspace{-1cm}\\
	\vspace{-.5cm}{\textbf{6B31}}
\\
	\includegraphics[scale=0.3]{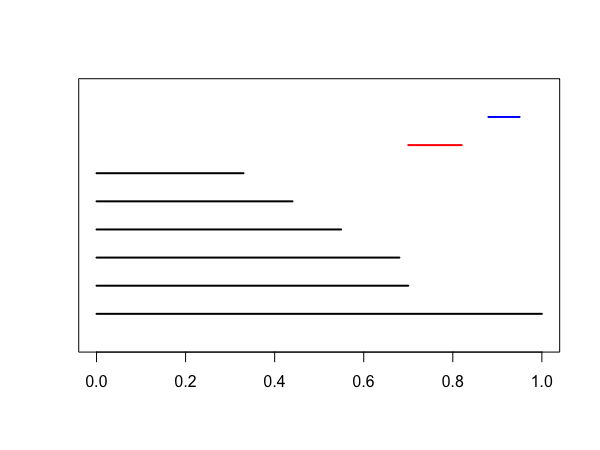} \hspace{-1cm}\\
	\vspace{-.5cm}{\textbf{6B32}}
\\
	\includegraphics[scale=0.3]{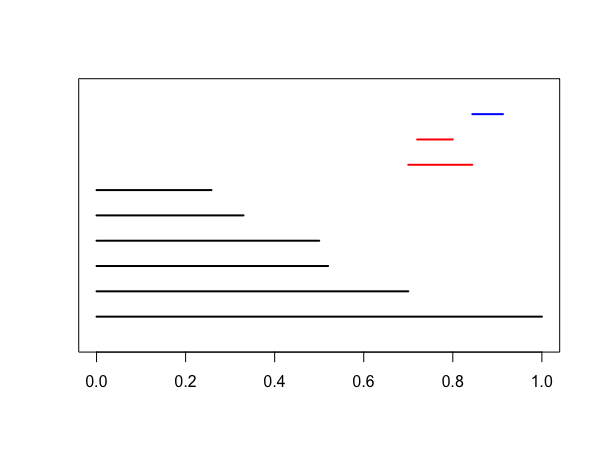} \hspace{-1cm}\\
	\vspace{-.5cm}{\textbf{6B33}}
\end{figure}
\FloatBarrier

\end{document}